\documentclass[twoside,twocolumn,9pt]{article}
\usepackage{extsizes}
\usepackage[super,sort&compress,comma]{natbib} 
\usepackage[version=3]{mhchem}
\usepackage[left=1.5cm, right=1.5cm, top=1.785cm, bottom=2.0cm]{geometry}
\usepackage{balance}
\usepackage{mathptmx}
\usepackage{sectsty}
\usepackage{graphicx} 
\usepackage{lastpage}
\usepackage[format=plain,justification=justified,singlelinecheck=false,font={stretch=1.125,small,sf},labelfont=bf,labelsep=space]{caption}
\usepackage{float}
\usepackage{fancyhdr}
\usepackage{fnpos}
\usepackage[english]{babel}
\addto{\captionsenglish}{%
  
}
\usepackage{array}
\usepackage{droidsans}
\usepackage{charter}
\usepackage[T1]{fontenc}
\usepackage[usenames,dvipsnames]{xcolor}
\usepackage{setspace}
\usepackage[compact]{titlesec}
\usepackage{hyperref}

\usepackage{epstopdf}

\definecolor{cream}{RGB}{222,217,201}

\begin{document}

\pagestyle{fancy}
\thispagestyle{plain}
\fancypagestyle{plain}{
\renewcommand{\headrulewidth}{0pt}
}

\makeFNbottom
\makeatletter
\renewcommand\LARGE{\@setfontsize\LARGE{15pt}{17}}
\renewcommand\Large{\@setfontsize\Large{12pt}{14}}
\renewcommand\large{\@setfontsize\large{10pt}{12}}
\renewcommand\footnotesize{\@setfontsize\footnotesize{7pt}{10}}
\makeatother

\renewcommand{\thefootnote}{\fnsymbol{footnote}}
\renewcommand\footnoterule{\vspace*{1pt}%
\color{cream}\hrule width 3.5in height 0.4pt \color{black}\vspace*{5pt}} 
\setcounter{secnumdepth}{5}

\makeatletter 
\renewcommand\@biblabel[1]{#1}            
\renewcommand\@makefntext[1]%
{\noindent\makebox[0pt][r]{\@thefnmark\,}#1}
\makeatother 
\renewcommand{\figurename}{\small{Fig.}~}
\sectionfont{\sffamily\Large}
\subsectionfont{\normalsize}
\subsubsectionfont{\bf}
\setstretch{1.125} 
\setlength{\skip\footins}{0.8cm}
\setlength{\footnotesep}{0.25cm}
\setlength{\jot}{10pt}
\titlespacing*{\section}{0pt}{4pt}{4pt}
\titlespacing*{\subsection}{0pt}{15pt}{1pt}

\fancyhead{}
\renewcommand{\headrulewidth}{0pt} 
\renewcommand{\footrulewidth}{0pt}
\setlength{\arrayrulewidth}{1pt}
\setlength{\columnsep}{6.5mm}
\setlength\bibsep{1pt}

\makeatletter 
\newlength{\figrulesep} 
\setlength{\figrulesep}{0.5\textfloatsep} 

\newcommand{\topfigrule}{\vspace*{-1pt}%
\noindent{\color{cream}\rule[-\figrulesep]{\columnwidth}{1.5pt}} }

\newcommand{\botfigrule}{\vspace*{-2pt}%
\noindent{\color{cream}\rule[\figrulesep]{\columnwidth}{1.5pt}} }

\newcommand{\dblfigrule}{\vspace*{-1pt}%
\noindent{\color{cream}\rule[-\figrulesep]{\textwidth}{1.5pt}} }

\makeatother

\twocolumn[
  \begin{@twocolumnfalse}
\vspace{1em}
\sffamily
\begin{tabular}{m{2cm} p{14cm} }

 & \noindent\LARGE{\textbf{Direct Neutron Detectors based on Carborane Containing Conjugated Polymers}} \\
\vspace{0.3cm} & \vspace{0.3cm} \\

 & \noindent\large{Aled Horner,\textit{$^{a}$} Fani E. Taifakou,\textit{$^{a}$} Choudhry Z. Amjad,\textit{$^{a}$} Filip Aniés,\textit{$^{b}$} Elizabeth George,\textit{$^{a}$} Chris Allwork,\textit{$^{a,c}$} Adrian J. Bevan,\textit{$^{a}$}$^{\ast}$ Martin Heeney,\textit{$^{b}$} and Theo Kreouzis\textit{$^{a}$}\textsuperscript{\textdagger}} \\
\vspace{0.3cm} & \vspace{0.3cm} \\
 & \noindent\normalsize{Thermal neutron detectors are crucial to a wide range of applications, including nuclear safety and security, cancer treatment, space research, non-destructive testing, and more. However, neutrons are notoriously difficult to capture due to their absence of charge, and only a handful of isotopes have a sufficient neutron cross-section. Meanwhile, commercially available $^{3}$He gas filled proportional counters suffer from depleting $^{3}$He feedstocks and complex device structures. In this work, we explore the potential of a carborane containing conjugated polymer (\textit{o}CbT$_{2}$-NDI) as a thermal neutron detector. The natural abundance of $^{10}$B in such a polymer enables intrinsic thermal neutron capture of the material, making it the first demonstration of an organic semiconductor with such capabilities. In addition, we show that thermal neutron detection can be achieved also by adding a $^{10}$B$_{4}$C sensitiser additive to the analogous carborane-free polymer PNDI(2OD)2T, whereas unsensitised PNDI(2OD)2T control devices only respond to the fast neutron component of the radiation field. This approach allows us to disentangle the fast and thermal neutron responses of the devices tested and compare the relative performance of the two approaches to thermal neutron detection. Both the carborane containing and the $^{10}$B$_{4}$C sensitised devices displayed enhancement due to thermal neutrons, above that of the unsensitised polymer. The detector response is found to be linear with flux up to $1.796$\,$\times$\,$10^7$\,cm$^{-2}$s$^{-1}$ n$_{th}\bar{v}$ and the response increases with drive bias, saturating at: ($56$\,$\pm$\,$3$)\,pA for \textit{o}CbT$_{2}$-NDI, ($59$\,$\pm$\,$9$)\,pA for $^{10}$B sensitised PNDI(2OD)2T, and ($36$\,$\pm$\,$3$)\,pA for PNDI(2OD)2T. This study demonstrates the viability of carboranyl polymers as neutron detectors, highlights the inherent chemical tuneability of organic semiconductors, and opens the possibility of their application to a number of different low-cost, scalable, and easily processable detector technologies.} \\

\end{tabular}

 \end{@twocolumnfalse} \vspace{0.6cm}

  ]

\renewcommand*\rmdefault{bch}\normalfont\upshape
\rmfamily
\section*{}
\vspace{-1cm}


\footnotetext{\textit{$^{a}$~Department of Physics and Astronomy, Queen Mary University of London, London, E1 4NS, United Kingdom.}}
\footnotetext{\textit{$^{b}$~Physical Science and Engineering Division, KAUST Solar Center (KSC), King Abdullah University of Science and Technology (KAUST), Thuwal, Saudi Arabia.}}
\footnotetext{\textit{$^{c}$~AWE, Aldermaston, Reading, RG7 4PR, United Kingdom.}}
\footnotetext{$\ast$~Corresponding Author e-mail: a.j.bevan@qmul.ac.uk.}
\footnotetext{\textdagger~Deceased.}


\section{Introduction}
\begin{figure*}
 \centering
 \includegraphics[height=10cm]{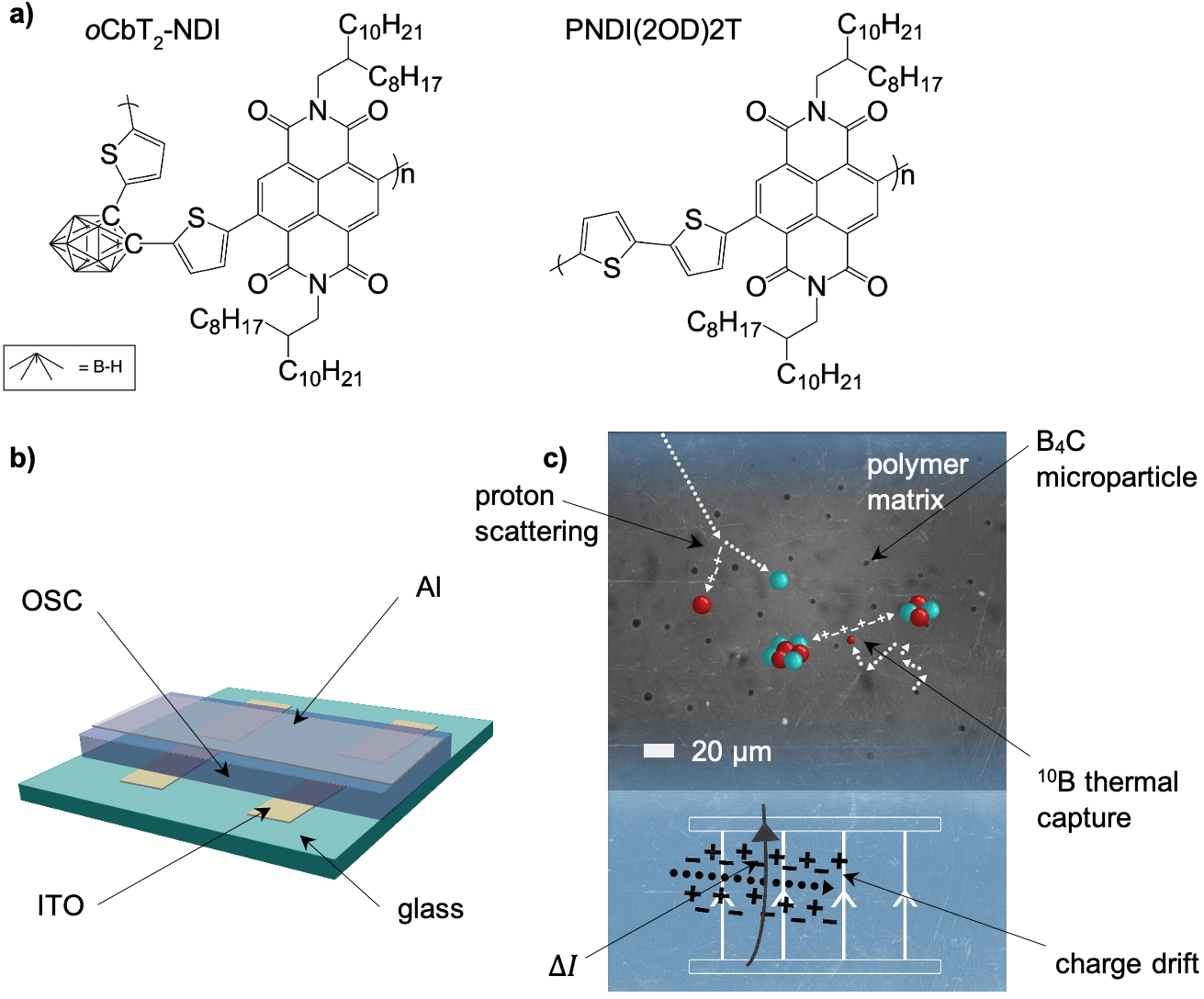}
 \caption{\textbf{a)} Chemical structure of the \textit{o}CbT$_{2}$-NDI and PNDI(2OD)2T polymers. \textbf{b)} Structure of devices fabricated on substrate. The overlap area between the ITO anodes and Al cathode defines an individual device. \textbf{c)} Top panel: optical reflection micrograph of a PNDI(2OD)2T:B$_4$C device showing B$_4$C microparticles in the form of dark spots. The fast and thermal neutron interactions are superimposed graphically and do not correspond in scale to the bar shown in the micrograph. Bottom panel: Schematic cross section through a single device. The current direction is from anode to cathode. Ionising (charged) neutron reaction products increase the charge carrier density within the device thus increasing the device current.}
 \label{fgr:Fig1}
\end{figure*}

Neutron detectors play a crucial part in a wide range of applications, including the nuclear industry, safeguarding radioactive material, neutron imaging, non-destructive testing, port of entry monitoring, studying space weather effects on commercial electronics, and for fundamental science in nuclear and particle physics. Imaging and non-destructive testing makes use of large area segmented neutron detectors. Monitoring of space weather effects and associated neutron damage on commercial electronics used in aviation is important to ensure equipment is retired before end-of-life failure of the electronics. Particle and nuclear physics experiments use neutron detectors to measure the properties of the neutron or for example to detect the presence of neutron background for low background searches of dark matter or studying the properties of neutrinos. In addition to these applications where detection of low fluxes or individual neutrons is important, there are also high flux detector demands for these scientific fields, for example at the European Spallation Source facility.\cite{ABELE20231}
Commercial thermal neutron detectors typically rely on $^{3}$He gas-filled proportional counters. However, $^{3}$He is primarily produced as a by-product of the nuclear industry, and depleting $^{3}$He stockpiles have created a need for non-$^{3}$He neutron detector technologies.\cite{KOUZES20101035,KOUZES2015172} Few isotopes exhibit a high thermal neutron absorption cross-section, and only $^{10}$B and $^{6}$Li are considered realistic alternatives to $^{3}$He. Alternatives include various Li- and B-based scintillators\cite{FAVALLI25,MAHL20181} and $^{10}$B-lined tubes.\cite{Weimar20} A major drawback of these solutions is that they rely on indirect detection methods, i.e. separate components for neutron absorption and subsequent light/charge detection. By employing a material which can both absorb neutrons and transport charge, device complexity may be reduced significantly. A direct-conversion detector has been demonstrated employing a 2D $^{6}$LiInP$_2$Se$_6$ semiconductor, however the fabrication of large area devices containing such semiconductor can be complex.\cite{CHICA20}
Detectors based on organic semiconductors (OSCs) are generally easier to fabricate as large-area films by a variety of solution or vacuum based processing. Organic electronic devices hence offer a scalable, tuneable, low-cost technology for detecting $\beta$\cite{SUZUKI2014304} and X-ray\cite{Nanayakkara} radiation as well as hadronic radiation such as $\alpha$ particles\cite{Taifakou:2021,Beckerle:2000}, protons\cite{FRATELLI21}, and neutrons.\cite{ZHAO21,Borowiec:2022,Bevan:2024,Chatzispyroglou:2020,ANIES24} The hydrocarbon nature of these materials means they can detect fast neutrons by design, and the use of single crystals of OSCs to this end has been reported.\cite{ZHAO21} In addition to the detection of fast neutrons, it is possible to add sensitisers to extend the response to include thermal neutrons\cite{Borowiec:2022,Bevan:2024,Chatzispyroglou:2020} OSCs are a similar density to human tissue thus this technology may be useful for medical physics, for example with boron neutron capture therapy dose monitoring and understanding neutron interactions within human tissue. The use of polymer semiconductors additionally allows for low temperature, large area solution-based fabrication methods.
We recently reported our preliminary work upon blends of a polymeric OSC with a $^{10}$B enriched sensitiser (B$_4$C).\cite{Borowiec:2022,Bevan:2024} The presence of the additive allows the fabrication of a solution processed device that could detect both fast and thermal neutrons. In that study we added $^{10}$B enriched B$_4$C sensitiser up to $20$\,\% by weight to devices fabricated using an n-type polymeric semiconductor (PNDI(2HD)2T). Whilst detection was observed, the device efficiency was significantly reduced from its theoretical maximum likely due to the formation of aggregates of the B$_4$C clusters throughout the film.  
Here we explore the potential for a carborane-containing conjugated polymer \textit{o}CbT$_{2}$-NDI for thermal neutron detection, as one solution for distributing the boron more evenly in a device. Conjugated carboranyl species have previously been demonstrated for conventional organic electronic applications, including light-emitting diodes, but have never before been demonstrated for neutron detection.\cite{ANIES24,FURUE16,MARSHALL13} Hence, we present the first example of an organic semiconductor intrinsic detection of thermal neutrons. \textit{o}CbT$_{2}$-NDI is compared to the closely related PNDI(2OD)2T, a copolymer of naphthalene diimide (NDI) and bithiophene (see Fig.~\ref{fgr:Fig1}a).\cite{ANIES2022124481} We test PNDI(2OD)2T devices with and without $^{10}$B enriched B$_4$C impurities to disentangle fast and thermal neutron contributions following our previously published approach. The level of $^{10}$B in the PNDI(2OD)2T devices is tuned to match the content in the \textit{o}CbT$_{2}$-NDI to allow direct comparison to the novel material.

\section{Organic semiconductor detectors}
The structures of both polymers are shown in Fig.~\ref{fgr:Fig1}a, and the only difference is the inclusion of an ortho-carborane between the two thiophene groups in \textit{o}CbT$_{2}$-NDI. This results in a significantly different shape for the two materials, with the presence of the carborane causing a significant backbone kink in an otherwise linear backbone, and resulting in a loss of crystallinity compared to PNDI(2OD)2T. The presence of the carborane also results in a reduction in molecular weight (see Supplementary Information (SI) Table S1), likely due to steric effects, although their different shape in solution makes it difficult to compare them directly by gel permeation chromatography.\cite{ANIES24} 
We fabricate devices on soda-lime glass as substrates (illustrated in Fig.~\ref{fgr:Fig1}b) using the methodology presented in the supplementary information (SI Section SI2).  These devices have indium-tin-oxide (ITO) anodes and aluminium cathodes, and are operated under ambient conditions. The active layer thickness, as prepared by drop-casting, varied between $\sim$$11$\,{\textmu}m and $\sim$$19$\,{\textmu}m (see SI Table S2). The standard individual device area is $4$\,mm$^2$ for all compositions, but we additionally produced a single $36$\,mm$^2$ \textit{o}CbT$_{2}$-NDI device.  In all, we fabricated \textit{o}CbT$_{2}$-NDI, B$_4$C sensitised PNDI(2OD)2T (denoted as PNDI(2OD)2T:B$_4$C), and pure PNDI(2OD)2T devices. The \textit{o}CbT$_{2}$-NDI contains natural abundance $^{10}$B in the carborane. This $^{10}$B content was approximately matched by dispersing $96.6$\,\% $^{10}$B isotope enriched B$_4$C into PNDI(2OD)2T solution, forming the PNDI(2OD)2T:B$_4$C. The B$_4$C content of the PNDI(2OD)2T:B$_4$C was approximately $2.75$\,wt\% (for detailed calculations see supplementary information section SI3). This enabled a direct comparison of the fast versus thermal neutron response of the PNDI(2OD)2T devices versus the \textit{o}CbT$_{2}$-NDI devices  under the same radiation field conditions.
Prior to neutron exposure, all devices were characterised by performing diagnostic $^{241}$Am $\alpha$ particle detection measurements. A noteworthy feature of these NDI polymer-based devices is that their radiation responses are stable in air over long periods of time (more than $800$ days). Additionally, following only a few days of conditioning in air, the dark currents drop to the order of tens of pA or lower when biased at voltages in the range explored here (up to tens of volts), then remaining stable. Thus, in contrast to our previous work with other OSCs \cite{Taifakou:2021, Borowiec:2022}, devices made from these materials improve after initial conditioning and remain stable for a time in excess of $1.3$ years. 

\section{Experimental setup}
As with our previous work\cite{Borowiec:2022}, we use the thermal pile at the UK’s National Physical Laboratory (NPL) in Teddington to provide a radiation field of neutrons.\cite{Kolkowski:1999,Hawkes:2018} Here, a Van de Graaff generator is used to accelerate a deuterium beam which impinges on Be targets, creating neutrons. The targets are surrounded by graphite to moderate the neutron energy, creating a field of predominately thermal neutrons ($0.025$ eV), as well as a higher energy tail of fast neutrons out to $\sim$$11$ MeV. Fluence rates provided by NPL during testing are written as n$_{th}\bar{v}$, where this corresponds to the sub-cadmium-cut-off ($0.5$ eV) fluence rate. The total field of neutrons is proportional to this given fluence rate. Any device exposed in the thermal pile will respond to a radiation field over a wide range of neutron energies. In the case of thermal neutrons, the sensitised devices interact via boron neutron capture (BNC) due to the inclusion of $^{10}$B, via the reaction n + $^{10}$B → $\alpha$ + $^7$Li (+$\gamma$) (see Fig.~\ref{fgr:Fig1}c). For higher energy neutrons the predominantly hydrocarbon nature of OSCs allows them to respond via elastic neutron-proton (n,p) scattering (Fig.~\ref{fgr:Fig1}c), as well as inelastic neutron-carbon (n,C) scattering above a certain energy threshold (see\cite{Borowiec:2022} for details). We note that OSCs are low density materials and that the devices have radiation lengths of less than $0.01$\,\% for $\gamma$ photons, allowing us to ignore any $\gamma$ component in our interpretation. Finally, the device response is based on electron-hole pair excitation across the OSC energy gap resulting from the passage of charged hadrons, i.e. p, $\alpha$, $^7$Li, C, through a biased device (see Figure~\ref{fgr:Fig1}c, lower panel).
We exposed devices in the access hole of the NPL thermal pile to neutron fluence rates ranging from $0.3$--$1.8$\,$\times$\,$10^7$\,cm$^{-2}$s$^{-1}$ n$_{th}\bar{v}$, corresponding to fluxes of $0.12$--$7.2$\,$\times$\,$10^5$ n/s in the $4$\,mm$^2$ devices. The neutron field is switched on and off by the insertion of a physical shutter in the deuterium beamline.  We define a signal response as a current enhancement ($\Delta I$), the observed change in current between beam off and beam on data acquisition periods. The current is measured using Keithley source measure units (2635B and 2470), that have been cross calibrated using a common reference device in the laboratory and have been verified to give the same response within instrument measurement accuracy. Pairs of devices were measured simultaneously as follows: PNDI(2OD)2T:B$_4$C together with pure PNDI(2OD)2T, and small area \textit{o}CbT$_{2}$-NDI with large area \textit{o}CbT$_{2}$-NDI. The temperature and relative humidity (RH) of the experimental hall were stable during data acquisition ($21$--$23$\textdegree{}C and $35$--$40$\,\%RH).

\section{Results}
\begin{figure*}
 \centering
 \includegraphics[height=10cm]{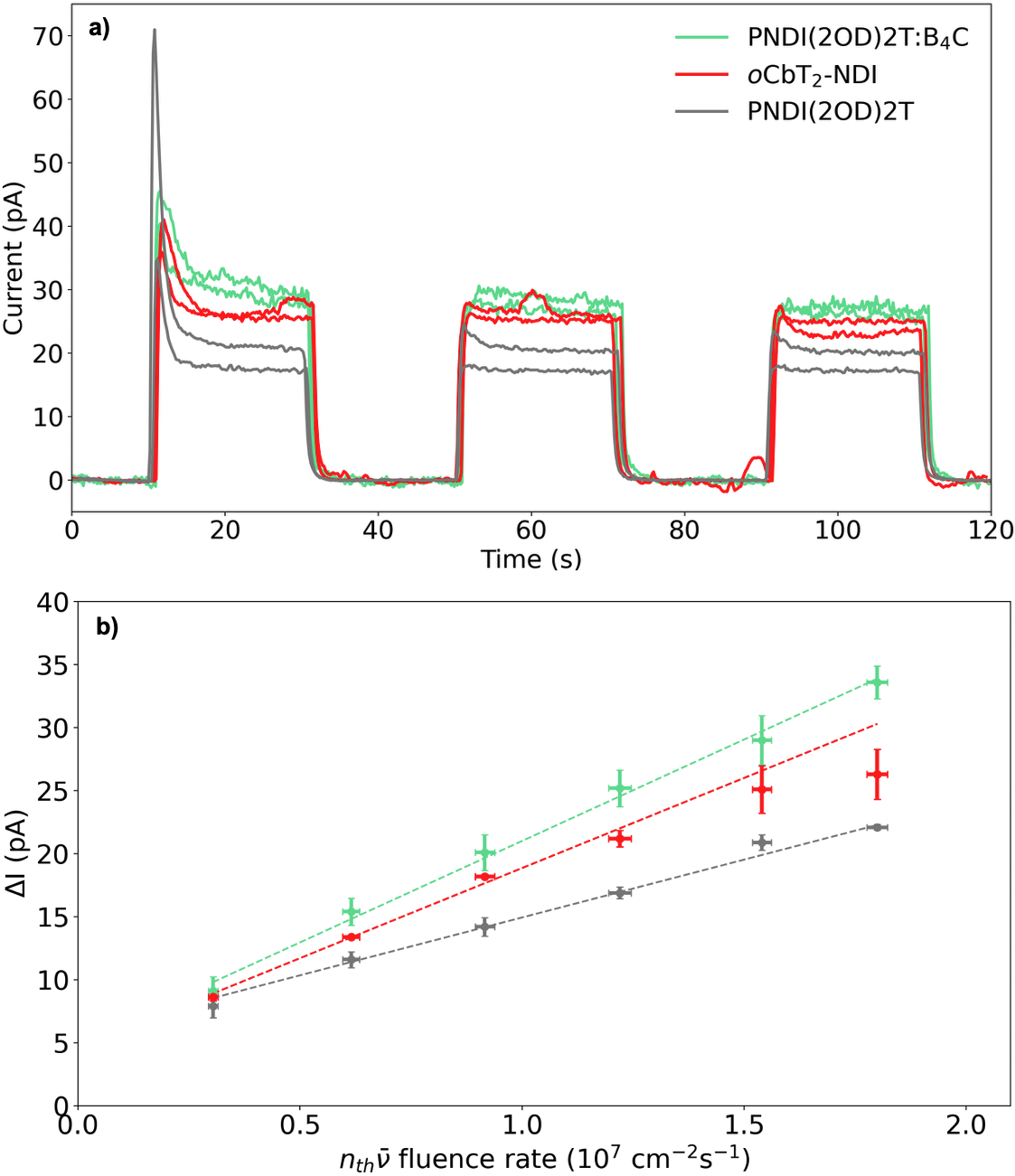}
 \caption{\textbf{a)} Device current versus time plots in the alternating absence and presence of neutrons at $+10$\,V bias. The current enhancement due to the $1.537$\,$\times$\,$10^7$\,cm$^{-2}$s$^{-1}$ n$_{th}\bar{v}$ neutron fluence rate is clearly distinguishable as a sharp rise on exposure, followed by a sharp drop when the neutron field is turned off. The figure shows responses from two individual $4$\,mm$^2$ devices for each of the following compositions: PNDI(2OD)2T:B$_4$C (green), \textit{o}CbT$_2$-NDI (red), and PNDI(2OD)2T (grey). \textbf{b)} The current enhancement ($\Delta I$) upon neutron exposure at $+10$\,V bias versus neutron fluence rate for a single device of each composition with linear fits to each data set. Both plots share the same legend shown in \textbf{a)}.}
 \label{fgr:Fig2}
\end{figure*}

By comparing the sensitised to the unsensitised device responses, we can separate the thermal neutron response from the higher energy neutron response of the OSC itself. We note that in a related study where elemental boron sensitised OSC devices were exposed to the same neutron field as in this study, the fast neutron OSC response itself was not isolated.\cite{Chatzispyroglou:2020} Fig.~\ref{fgr:Fig2}a shows the response of our devices ($4$\,mm$^2$) when exposed to neutrons at $+10$\,V bias and fluence rate $1.537$\,$\times$\,$10^7$\,cm$^{-2}$s$^{-1}$ n$_{th}\bar{v}$. We note that each time a measurement run starts there is a short period where feedback from beam monitors results in a beam current adjustment. This can be seen in Fig.~\ref{fgr:Fig2}a as the spike at the start of a given set of beam on/off measurements. All plots have had the leakage current background subtracted using the previously described method.\cite{Taifakou:2021} The PNDI(2OD)2T:B$_4$C and \textit{o}CbT$_{2}$-NDI devices show enhanced response to the neutron field compared to the PNDI(2OD)2T devices. The average current enhancement, $\Delta I$, for each type of device under neutron irradiation at $+10$\,V bias were  ($28$\,$\pm$\,$1$)\,pA  for PNDI(2OD)2T:B$_4$C, ($25$\,$\pm$\,$1$)\,pA for \textit{o}CbT$_{2}$-NDI, and ($19$\,$\pm$\,$1.5$)\,pA for PNDI(2OD)2T. All devices interact with the fast neutron component of the radiation field, and we interpret the increased current enhancement in PNDI(2OD)2T:B$_4$C and \textit{o}CbT$_{2}$-NDI devices as being due to the BNC interaction. By comparing the sensitised device responses to the PNDI(2OD)2T responses we note that the sensitised samples give a current enhancement $\sim$$40$\,\% larger than that of the purely organic device. Henceforth we denote this increase due to BNC as the signal enhancement. With only $3\sigma$ difference, we see good agreement between the PNDI(2OD)2T:B$_4$C and \textit{o}CbT$_{2}$-NDI signal enhancements. This suggests that there is no significant aggregation of the B$_4$C component of the binary mixture, since that would be expected to cause a reduction in the thermal neutron detection efficiency. Optical micrographs (Fig.~\ref{fgr:Fig1}c) of the films support this observation, where the average B$_4$C particle size is $6$\,{\textmu}m (not too significantly above the supplier rating of $<5$\,{\textmu}m).

Fig.~\ref{fgr:Fig2}b shows current enhancement upon neutron exposure at $+10$\,V bias versus neutron fluence rate of ($0.3$--$1.8$)\,$\times$\,$10^7$\,cm$^{-2}$s$^{-1}$ n$_{th}\bar{v}$ for the three types of $4$\,mm$^2$ devices.  All device responses are linear with neutron flux up to the maximum value the accelerator can deliver, consistent with findings of our previous work using PNDI(2HD)2T.\cite{Borowiec:2022} The signal enhancement due to thermal neutron capture in the PNDI(2OD)2T:B$_4$C and \textit{o}CbT$_{2}$-NDI devices compared to the PNDI(2OD)2T only device is clearly evident, with the linear fit gradients for each type of device being: ($1.6$\,$\pm$\,$0.1$)\,$\times$\,$10^{-6}$\,pAcm$^2$s for PNDI(2OD)2T:B$_4$C, ($1.43$\,$\pm$\,$0.08$)\,$\times$\,$10^{-6}$\,pAcm$^2$s for \textit{o}CbT$_{2}$-NDI, and ($0.92$\,$\pm$\,$0.04$)\,$\times$\,$10^{-6}$\,pAcm$^2$s for PNDI(2OD)2T. Using these gradients the signal enhancement may be calculated as $66$\,\% for the samples that contain boron compared with the purely organic samples, with less than $2\sigma$ difference between these results.

\begin{figure*}
 \centering
 \includegraphics[height=10cm]{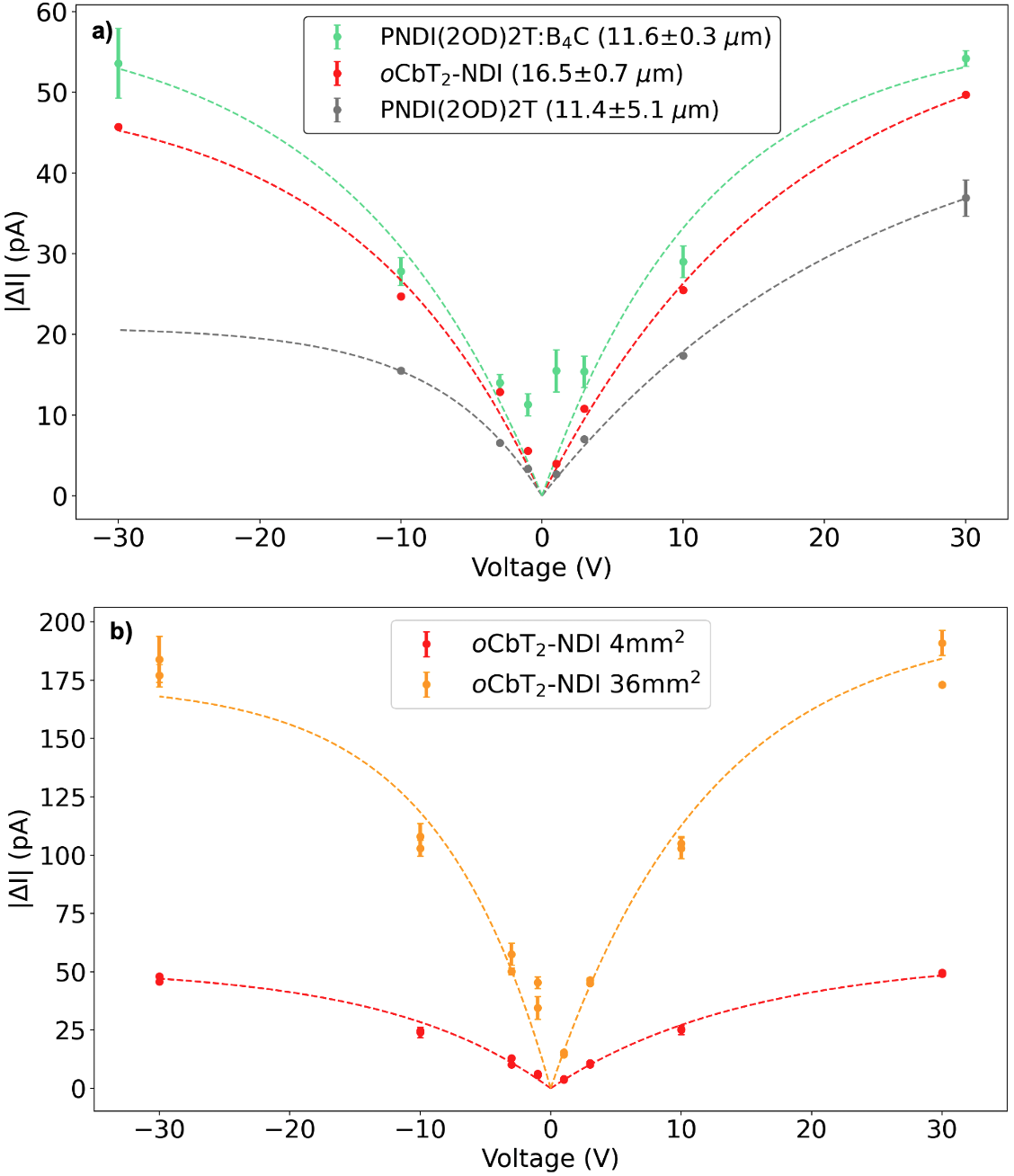}
 \caption{\textbf{a)} Current enhancement versus bias characteristics for PNDI(2OD)2T:B$_4$C (green), \textit{o}CbT$_2$-NDI (red), and PNDI(2OD)2T (grey) $4$\,mm$^2$ devices of similar thickness. \textbf{b)} Current enhancement versus bias characteristics for $36$\,mm$^2$ \textit{o}CbT$_2$-NDI (orange) and $4$\,mm$^2$ \textit{o}CbT$_2$-NDI (red). All data sets were obtained at a fluence rate of $1.537$\,$\times$\,$10^7$\,cm$^{-2}$s$^{-1}$ n$_{th}\bar{v}$. The dashed lines are fits to Eq.~\ref{eq:DI} for each data set in both figures.}
 \label{fgr:Fig3}
\end{figure*}

Fig.~\ref{fgr:Fig3}a shows the current enhancement versus bias characteristics for PNDI(2OD)2T:B$_4$C, \textit{o}CbT$_{2}$-NDI, and PNDI(2OD)2T $4$\,mm$^2$ devices of similar thickness under constant neutron field conditions. Fig.~\ref{fgr:Fig3}b shows the comparison for the $36$\,mm$^2$ and $4$\,mm$^2$ \textit{o}CbT$_{2}$-NDI devices. The responses are approximately symmetric and increase with increasing bias, although there is evidence of saturation in the response, as shown by the curvature of the plots. We attribute the increase in response to improved charge extraction under higher bias, and the saturation to the devices approaching the maximum possible charge extracted. The enhancement-bias characteristics can be fitted using the simple exponential approach given by

\begin{equation}\label{eq:DI}
    \Delta I = I_{\text{sat}} \left(1-e^{-\left(V/V_0 \right)}\right).
\end{equation}

Here $\Delta I$ is the current enhancement, $V$ is the device bias, $I_{\text{sat}}$ is the asymptotic limit of $\Delta I$, and $V_0$ is the characteristic bias indicating the graph curvature. The $I_{\text{sat}}$ and $V_0$ fitting parameters obtained for the different types of devices are summarised in the supplementary information Table S2. The response enhancement due to the boron sensitisation in the PNDI(2OD)2T:B$_4$C and \textit{o}CbT$_{2}$-NDI devices compared to the PNDI(2OD)2T shown in Figs.~\ref{fgr:Fig2}a and~\ref{fgr:Fig2}b is also evident in Fig.~\ref{fgr:Fig3}a. The response increases with voltage bias, saturating at: ($59$\,$\pm$\,$9$)\,pA for PNDI(2OD)2T:B$_4$C, ($56$\,$\pm$\,$3$)\,pA for \textit{o}CbT$_{2}$-NDI, and ($36$\,$\pm$\,$3$)\,pA for PNDI(2OD)2T at a fluence rate of $1.537$\,$\times$\,$10^7$\,cm$^{-2}$s$^{-1}$ n$_{th}\bar{v}$. Thus boron sensitisation gives a signal enhancement of $60$\,\% when comparing these saturation currents.

We additionally note that the individual device response to the neutron field scales with increasing area (Fig.~\ref{fgr:Fig3}b), demonstrating the suitability of organic neutron detectors for large-area devices. We note that the observed increase of a factor of $3.5$\,$\pm$\,$0.3$ is less than the factor of 9 difference in device area between the $4$\,mm$^2$ and $36$\,mm$^2$ \textit{o}CbT$_{2}$-NDI devices. This factor has been evaluated using the saturation currents obtained by fits to Eq.~\ref{eq:DI}, namely ($56$\,$\pm$\,$3$)\,pA for the $4$\,mm$^2$ device and ($196$\,$\pm$\,$10$)\,pA for the $36$\,mm$^2$ device. This discrepancy may be due to a non-proportional series resistance within the larger area device leading to less than proportional increases in current enhancement with the higher area \cite{Korde:2003,Scholze:2004}. Additional $\alpha$ particle detection measurements carried out on the same devices using the saturation currents gave this area response factor as $2.1$\,$\pm$\,$0.2$, also lower than the expected factor of 9. When instead evaluated at a given bias for neutron irradiation ($+10$\,V data, as seen in Fig.~\ref{fgr:Fig3}b), this factor is $4.1$\,$\pm$\,$0.1$, and the corresponding $\alpha$ irradiation value is $4.8$\,$\pm$\,$0.1$.  These larger factors imply a better device efficiency at lower biases than at biases corresponding to the saturation current. Using either method, we do not obtain the expected nine-fold increase, indicating a non-linear relationship between device area and responsivity. The increased roughness ($18.9$\,$\pm$\,$3.1$\,{\textmu}m thickness) of the large area device, compared to the small area device ($16.5$\,$\pm$\,$0.7$\,{\textmu}m thickness) is another factor to consider in this line of research.
To assess the efficiency of the thermal neutron capture process, we compare the calculated theoretical upper limit for the capture quantum efficiency (QE) with an empirical thermal neutron conversion efficiency ($\eta_{\text{n conv}}$). A larger value of the experimentally found $\eta_{\text{n conv}}$  which is closer to the theoretical maximum QE is indicative of a more efficient device. Hence it is desirable to maximise the value of ($\eta_{\text{n conv}}$/QE )\,\%. The QE is obtained using the number density of the $^{10}$B content and the thermal neutron capture cross section. The value of $\eta_{\text{n conv}}$ is obtained using thermal neutron irradiation data compared with $\alpha$ particle irradiation characterising data. Detailed calculations for both QE and $\eta_{\text{n conv}}$  are given in the supplementary information (Section SI4). The QE of all $4$\,mm$^2$ sized PNDI(2OD)2T:B$_4$C was calculated as $0.89$\,\%. The corresponding value for \textit{o}CbT$_{2}$-NDI was estimated to be around $0.69$\,\%. We obtain $\eta_{\text{n conv}}$ values of $\sim$$0.20$\,\% and $\sim$$0.14$\,\% for PNDI(2OD)2T:B$_4$C and \textit{o}CbT$_{2}$-NDI devices respectively at $+10$\,V bias. 
Our previous work with poor B$_4$C dispersion resulted in an ($\eta_{\text{n conv}}$/QE)\,\% value of $\sim$$11$\,\%.\cite{Borowiec:2022} In comparison here we find PNDI(2OD)2T:B$_4$C and \textit{o}CbT$_{2}$-NDI devices at $+10$\,V drive conditions give values of $\sim$$22$\,\%  and $\sim$$20$\,\% respectively, both an improvement on the $\sim$$11$\,\% fraction previously reported. The PNDI(2OD)2T polymer used in the present study has significantly improved solubility during processing, compared to the PNDI(2HD)2T used in our previous study. This may have aided improved dispersion of the B$_4$C component. Additionally, the lower B$_4$C loading in the present study is another factor that may contribute to this difference in sensitiser dispersion. 
There are routes to increasing the thermal neutron conversion efficiency in these types of devices. Ongoing studies using techniques such as independently dispersing B$_4$C in solvents as a colloidal solution before mixing with the organic material solution yields higher ($>20$\,wt\%) loadings with significantly improved dispersion in OSC films, as opposed to the previous procedure of dry mixing B$_4$C with the organic material followed by solution processing. This order of magnitude increase in B$_4$C content may significantly improve the current enhancement by enabling higher doping. In the case of \textit{o}CbT$_{2}$-NDI, $^{10}$B isotope enriched carborane can be used (immediately offering a fivefold increase in efficiency over natural abundance) and we note that the carborane does not necessarily need to only be included in the polymer backbone; it could also be attached to the alkane chains, increasing the numbers of carboranes included per monomer and thus the overall $^{10}$B concentration. If both these approaches were realised (and incorporating these side chain carboranes did not negatively affect solubility or electronic properties), then over an order of magnitude increase in efficiency could be achieved using CbT$_{2}$-NDI polymers. The blending of B$_4$C with CbT$_{2}$-NDI may also be an interesting approach, where the amorphous nature of the CbT$_{2}$-NDI polymer may help to suppress phase segregation with the B$_4$C. We further note that both PNDI(2OD)2T:B$_4$C and \textit{o}CbT$_{2}$-NDI device thickness could be increased to further enhance the thermal neutron detection efficiency.
There is some indication that the B$_4$C inclusion may have led to increased noise in dark current, compared to polymer only ($4$\,mm$^2$) devices. The standard deviation of the dark current noise are $2.1$\,pA for PNDI(2OD)2T:B$_4$C, $0.6$\,pA for \textit{o}CbT$_{2}$-NDI, and $0.3$\,pA for PNDI(2OD)2T. This phenomenon is initially counterintuitive, since the B$_4$C bandgap greatly exceeds the energy gap of the OSC, and thus the B$_4$C inclusions should not affect the electron and hole transport within the semiconductor from an electronic perspective. However, the inclusion of B$_4$C may disrupt the microstructure of the semicrystalline PNDI(2OD)2T polymer, which influence the energetics of the polymer as well as the charge transport characteristics.
We also note that the linear fits to the response versus neutron flux data (Fig.~\ref{fgr:Fig2}b) do not appear to go through the origin. This implies that the device response is greater than that due solely to the neutron flux. This would be consistent with the devices responding to short-lived secondary radiation resulting from neutron activation of the metal parts used in the detector assembly.  Radiation from material activation is dominated by the metal BNC connectors used on the plastic housing used to contain the devices and perform these measurements in a light-tight environment.
In conclusion, we demonstrate that diode devices fabricated from boron containing conjugated polymer films are able to detect both thermal and fast neutrons. This is the first example of such a detector where the boron component forms a significant part of the polymer backbone, rather than as a sensitizer added to the film. By the measurement of  PNDI(2OD)2T devices with and without $^{10}$B-enriched B$_4$C, we demonstrate that the thermal neutron response can be decoupled from the response to higher-energy fast neutrons, and that the former correlates quantitatively with the degree of $^{10}$B sensitization. Furthermore, we have close agreement between the calculated QE and measured $\eta_{\text{n conv}}$, both of the same order, obtained from sensitiser concentrations of $2.75$\,wt\%. We additionally observe a dark current noise increase for the PNDI(2OD)2T:B$_4$C devices compared to those without the additive. The \textit{o}CbT$_{2}$-NDI OSCs can be used as a thermal neutron detector, and we observe that signals from devices using this have lower noise than the PNDI(2OD)2T:B$_4$C alternative presented. Where signal to noise is paramount over quantum efficiency, the \textit{o}CbT$_{2}$-NDI devices are superior to the PNDI(2OD)2T based ones. Overall, this work demonstrates that carborane inclusion is an effective strategy to obtain thermal neutron capabilities of organic semiconductors. Future advances in material design may realise higher quantum efficiencies, e.g. higher $^{10}$B content through the inclusion of pendant carboranyl side chains and/or $^{10}$B enriched carborane. The detectors presented here use the same fabrication process used in commercially available large area organic electronic systems, indicating the potential for a semiconductor neutron detector technology that is scalable, and the viability of carborane-containing organic molecules for this neutron detection technique.

\section*{Acknowledgements}
The authors gratefully acknowledge the generous contributions of the following UK organisations, AWE Plc., Queen Mary University of London, and the Science and Technology Facilities Council (grant numbers ST/S00095X/1, ST/T002212/1, ST/V000039/1 and ST/T002212/1) to support this work. Furthermore we acknowledge the staff at NPL as well as the Technical Staff within the School of Physical and Chemical Sciences at Queen Mary University of London for technical support and assistance in completing this work. Copyright 2025 UK Ministry of Defence © Crown Owned Copyright 2025/AWE.

\section*{Author contributions}
Paper Writing: CZA, FA, AJB, AH, MH, TK, FET; Chemical Synthesis: FA, MH; Device Fabrication: AH, FET; Lab and Field Tests: CZA, AJB, EG, AH, TK, FET; Analysis: AH, FET; Interpretation: AJB, AH, TK, FET; Project oversight: CA, AJB, TK.

\section*{Conflicts of interest}
There are no conflicts to declare.

\section*{Data availability}
All relevant data are provided with the paper and its SI. 



\balance


\bibliography{rsc} 
\bibliographystyle{rsc} 

\pagebreak
\setcounter{section}{0}

\twocolumn[
  \begin{@twocolumnfalse}
\vspace{1em}
\sffamily
\begin{tabular}{m{2cm} p{14cm} }

 & \noindent\LARGE{\textbf{Direct Neutron Detectors based on Carborane Containing Conjugated Polymers: Supplementary Information}} \\
\vspace{0.3cm} & \vspace{0.3cm} \\

 & \noindent\large{Aled Horner,\textit{$^{a}$} Fani E. Taifakou,\textit{$^{a}$} Choudhry Z. Amjad,\textit{$^{a}$} Filip Aniés,\textit{$^{b}$} Elizabeth George,\textit{$^{a}$} Chris Allwork,\textit{$^{a,c}$} Adrian J. Bevan,\textit{$^{a}$}$^{\ast}$ Martin Heeney,\textit{$^{b}$} and Theo Kreouzis\textit{$^{a}$}\textsuperscript{\textdagger}} \\

\vspace{0.3cm} & \vspace{0.3cm} \\
 & \noindent\normalsize{Thermal neutron detectors are crucial to a wide range of applications, including nuclear safety and security, cancer treatment, space research, non-destructive testing, and more. However, neutrons are notoriously difficult to capture due to their absence of charge, and only a handful of isotopes have a sufficient neutron cross-section. Meanwhile, commercially available $^{3}$He gas filled proportional counters suffer from depleting $^{3}$He feedstocks and complex device structures. In this work, we explore the potential of a carborane containing conjugated polymer (\textit{o}CbT$_{2}$-NDI) as a thermal neutron detector. The natural abundance of $^{10}$B in such a polymer enables intrinsic thermal neutron capture of the material, making it the first demonstration of an organic semiconductor with such capabilities. In addition, we show that thermal neutron detection can be achieved also by adding a $^{10}$B$_{4}$C sensitiser additive to the analogous carborane-free polymer PNDI(2OD)2T, whereas unsensitised PNDI(2OD)2T control devices only respond to the fast neutron component of the radiation field. This approach allows us to disentangle the fast and thermal neutron responses of the devices tested and compare the relative performance of the two approaches to thermal neutron detection. Both the carborane containing and the $^{10}$B$_{4}$C sensitised devices displayed enhancement due to thermal neutrons, above that of the unsensitised polymer. The detector response is found to be linear with flux up to $1.796$\,$\times$\,$10^7$\,cm$^{-2}$s$^{-1}$ n$_{th}\bar{v}$ and the response increases with drive bias, saturating at: ($56$\,$\pm$\,$3$)\,pA for \textit{o}CbT$_{2}$-NDI, ($59$\,$\pm$\,$9$)\,pA for $^{10}$B sensitised PNDI(2OD)2T, and ($36$\,$\pm$\,$3$)\,pA for PNDI(2OD)2T. This study demonstrates the viability of carboranyl polymers as neutron detectors, highlights the inherent chemical tuneability of organic semiconductors, and opens the possibility of their application to a number of different low-cost, scalable, and easily processable detector technologies.} \\

\end{tabular}

 \end{@twocolumnfalse} \vspace{0.6cm}

  ]

\renewcommand*\rmdefault{bch}\normalfont\upshape
\rmfamily
\section*{}
\vspace{-1cm}


\footnotetext{\textit{$^{a}$~Department of Physics and Astronomy, Queen Mary University of London, London, E1 4NS, United Kingdom.}}
\footnotetext{\textit{$^{b}$~Physical Science and Engineering Division, KAUST Solar Center (KSC), King Abdullah University of Science and Technology (KAUST), Thuwal, Saudi Arabia.}}
\footnotetext{\textit{$^{c}$~AWE, Aldermaston, Reading, RG7 4PR, United Kingdom.}}
\footnotetext{$\ast$~Corresponding Author e-mail: a.j.bevan@qmul.ac.uk.}
\footnotetext{\textdagger~Deceased.}


\renewcommand{\thefigure}{SI\arabic{figure}}
\setcounter{figure}{0}
\renewcommand{\thetable}{S\arabic{table}}
\setcounter{table}{0}
\renewcommand{\theequation}{S\arabic{equation}}
\setcounter{equation}{0}
\renewcommand{\thesection}{SI\arabic{section}}    

\section{Polymer definitions}
\label{sec:PolyDef}
The polymers discussed in this paper are as follow:

\textbf{\textit{o}CbT$_2$-NDI}      poly\{[\textit{N},\textit{N}'-bis(2-octyldodecyl)naphthalene-1,4,5,8-bis(dicarboximide)-2,6-diyl]-\textit{alt}-5,5'-[1,12/7/2-bis(5-thiophen-2-yl)-\textit{ortho}-carborane]\};

\textbf{PNDI(2HD)2T}        poly\{[\textit{N},\textit{N}'-bis(2-hexyldecyl)naphthalene-1,4,5,8-bis(dicarboximide)-2,6-diyl]-\textit{alt}-5,5'-(2,2'-bithiophene)\};

\textbf{PNDI(2OD)2T}        poly\{[\textit{N},\textit{N}'-bis(2-octyldodecyl)naphthalene-1,4,5,8-bis(dicarboximide)-2,6-diyl]-\textit{alt}-5,5'-(2,2'-bithiophene)\}.

\section{Sample fabrication}
\label{sec:SampFab}
Samples are fabricated similar to our previous work described in the Supplementary Information of Borowiec et al.\cite{Borowiec:2022} Unlike in this previous work where PNDI(2HD)2T was used, here we instead use PNDI(2OD)2T from Ossila Ltd. (batch number M1201A3) and \textit{o}CbT$_2$-NDI from the authors of Aniés et al.\cite{ANIES2022124481} Relevant polymer weights and energy levels are given in Table S1. 

\begin{table*}[h]
\small
  \caption{Polymer weights, polydispersity index, and energy levels for the two organic materials used}
  \label{tbl:matdetails}
  \begin{tabular*}{\textwidth}{@{\extracolsep{\fill}}ccccccc}
    \hline
    \textbf{Material}  & \textbf{M$_{\text{n}}$ {[}Da{]}} & \textbf{M$_{\text{w}}$ {[}Da{]}} & \textbf{PDI} & \textbf{HOMO {[}eV{]}} & \textbf{LUMO {[}eV{]}} & \textbf{Energy Gap {[}eV{]}} \\
    \hline
    \textit{o}CbT$_2$-NDI & $5\,000$ & $9\,000$ & $1.73$ & $-6.70$ & $-3.98$ & $2.72$ \\
    PNDI(2OD)2T & $90\,982$ & $202\,261$ & $2.22$ & $-5.77$ & $-3.84$ & $1.93$ \\
    \hline
  \end{tabular*}
\end{table*}

ITO of thickness $150$\,nm is etched into its desired pattern on $400$\,mm$^2$ square soda lime glass substrates for each device type, forming bottom electrodes. For each set of $4$\,mm$^2$ device the ITO is etched via photolithography into a pattern consisting of four rectangles of ($2$\,$\times$\,$7$)\,mm. For the $36$\,mm$^2$ device, the ITO is etched leaving a ($6$\,$\times$\,$20$)\,mm strip on the substrate. The organic material (mixed with B$_4$C in the case of the sample requiring sensitisation) is dissolved in chloroform (nominally $8$\,mg/$0.5$\,mL) and stirred at $500$\,rpm for two hours at $80$\,\textdegree{}C. The solution is then drop cast onto the ITO-etched glass substrate at ambient temperature and left for $15$\,mins before returned to $80$\,\textdegree{}C hotplate for two hours to drive solvent evaporation, leaving dry films of the organic material. Dissolving, drop casting, and drying is all done in a nitrogen-rich controlled environment within a glovebox. Finally, a ($6$\,$\times$\,$20$)\,mm Al top electrode is applied through a shadow mask using vacuum deposition perpendicular to the bottom ITO electrodes. The overlapping regions of ITO and Al define the device regions on the substrates. Resulting device thicknesses are outlined in Table~\ref{tbl:thickandres}.

\begin{table*}[h]
\small
  \caption{Summary of device thicknesses and fitted parameters obtained using Eq.~\ref{eq:ndensity} for devices irradiated with neutrons at NPL. For ease of calculation, all fitting was done using the modulus of drive bias and response current hence the positive values in the table}
  \label{tbl:thickandres}
  \begin{tabular*}{\textwidth}{@{\extracolsep{\fill}}ccccccc}
    \hline
    \textbf{Sample name} & \textbf{\begin{tabular}[c]{@{}c@{}}Material \\ (size)\end{tabular}} & \textbf{Device number} & \textbf{Device thickness {[}{\textmu}m{]}} & \textbf{Drive bias} & \textbf{$I_{\text{sat}}$ {[}pA{]}} & \textbf{$V_0$ {[}V{]}} \\
    \hline
    AH008  & \begin{tabular}[c]{@{}c@{}}\textit{o}CbT$_2$-NDI \\ ($4$\,mm$^{2}$)\end{tabular}  & $4$ & $16.5$\,$\pm$\,$0.7$ & Negative & $51$\,$\pm$\,$6$ & $13.3$\,$\pm$\,$3.4$ \\
    AH008  & \begin{tabular}[c]{@{}c@{}}\textit{o}CbT$_2$-NDI \\ ($4$\,mm$^{2}$)\end{tabular}  & $4$ & $16.5$\,$\pm$\,$0.7$ & Positive & $60$\,$\pm$\,$4$ & $17.5$\,$\pm$\,$2.1$ \\
    AH018  & \begin{tabular}[c]{@{}c@{}}\textit{o}CbT$_2$-NDI \\ ($16$\,mm$^2$)\end{tabular} & $1$ & $18.9$\,$\pm$\,$3.1$ & Negative & $189$\,$\pm$\,$31$ & $9.8$\,$\pm$\,$4.1$ \\
    AH018  & \begin{tabular}[c]{@{}c@{}}\textit{o}CbT$_2$-NDI \\ ($16$\,mm$^2$)\end{tabular} & $1$ & $18.9$\,$\pm$\,$3.1$ & Positive & $219$\,$\pm$\,$10$ & $14.7$\,$\pm$\,$1.4$ \\
    AH018  & \begin{tabular}[c]{@{}c@{}}\textit{o}CbT$_2$-NDI \\ ($16$\,mm$^2$)\end{tabular} & $1$ & $18.9$\,$\pm$\,$3.1$ & Negative & $187$\,$\pm$\,$24$ & $11.1$\,$\pm$\,$3.5$ \\
    AH018  & \begin{tabular}[c]{@{}c@{}}\textit{o}CbT$_2$-NDI \\ ($16$\,mm$^2$)\end{tabular} & $1$ & $18.9$\,$\pm$\,$3.1$ & Positive & $188$\,$\pm$\,$7$  & $12.0$\,$\pm$\,$1.1$ \\
    AH019  & \begin{tabular}[c]{@{}c@{}}PNDI(2OD)2T:B$_4$C \\ ($4$\,mm$^{2}$)\end{tabular} & $4$ & $11.6$\,$\pm$\,$0.3$ & Negative & $60$\,$\pm$\,$12$  & $13.8$\,$\pm$\,$6.3$ \\
    AH019  & \begin{tabular}[c]{@{}c@{}}PNDI(2OD)2T:B$_4$C \\ ($4$\,mm$^{2}$)\end{tabular} & $4$ & $11.6$\,$\pm$\,$0.3$ & Positive & $57$\,$\pm$\,$14$  & $11.6$\,$\pm$\,$6.9$ \\
    AH020  & \begin{tabular}[c]{@{}c@{}}PNDI(2OD)2T \\ ($4$\,mm$^{2}$)\end{tabular} & $1$ & $1.4$\,$\pm$\,$5.1$ & Negative & $21$\,$\pm$\,$4$ & $7.4$\,$\pm$\,$2.6$ \\
    AH020  & \begin{tabular}[c]{@{}c@{}}PNDI(2OD)2T \\ ($4$\,mm$^{2}$)\end{tabular} & $1$ & $11.4$\,$\pm$\,$5.1$ & Positive & $50$\,$\pm$\,$4$ & $22.7$\,$\pm$\,$3.0$ \\
    \hline
  \end{tabular*}
\end{table*}

\section{Calculation of the $^{10}$B content in the \textit{o}CbT$_2$-NDI}
\label{sec:Bcont}
The density, $\rho$, of \textit{o}CbT$_2$-NDI is estimated from typical values from literature for comparable organic materials to range from $0.9$ to $1.2$\,g/cc. One polyhedral cluster of carborane contains $10$ boron atoms. There is one carborane moiety per monomer hence the number of boron atoms per monomer unit is also $10$. The total boron number density $n_{\text{B}}$ is therefore,

\begin{equation}
    n_\text{B} = 10 \frac{N_{\text{A}}}{M} \rho.
    \label{eq:ndensity}
\end{equation}

Where $N_{\text{A}}$ is Avogadro’s number and $M$ is the \textit{o}CbT$_2$-NDI molar mass per monomer unit ($1131.72$\,g/mol). This gives a range of values for $n_{\text{B}}$ from $4.8$ to $6.4$\,$\times$\,$10^{21}$\,cc$^{-1}$. Given the natural abundance of $^{10}$B is $19.9$\,\% of B atoms, this gives a $n_{10\text{B}}$ value ranging from $0.9$ to $1.3$\,$\times$\,$10^{21}$\,cc$^{-1}$ for \textit{o}CbT$_2$-NDI. To estimate the amount of $96.6$\,\% $^{10}$B enriched B$_4$C needed to match $^{10}$B content of a PNDI(2OD)2T:B$_4$C sample to that of an \textit{o}CbT$_2$-NDI sample, this value must be multiplied by the desired volume and molar density of B$_4$C ($55.255$\,g/mol) and divided by the enrichment percentage, $4$ (the number of B in B$_4$C), and Avogadro’s number. This gives the mass needed as ranging from $0.15$ to $0.19$\,mg. Given that both bounds of this range round to $0.2$\,mg, this value was taken for measuring out when fabricating these detectors. The true amount measured out was $0.22$\,mg, giving these detectors a $2.75$\,wt\% B$_4$C content.

\section{Quantum efficiency and thermal neutron conversion efficiency}
\label{sec:QEandeta}
The detector’s thermal neutron quantum efficiency (QE) is defined by the probability of thermal neutrons interacting with the detector such that

\begin{equation}
    \text{QE} = 1 - e^{-\frac{d}{\lambda}},
    \label{eq:QE1}
\end{equation}

where $d$ is device thickness and $\lambda= 1/(n_{10\text{B}}\sigma)$, where $\sigma$ is the interaction cross section. We can rewrite this as

\begin{equation}
    \text{QE} = 1 - e^{-\frac{N_{10\text{B}}\sigma}{A}},
    \label{eq:QE2}
\end{equation}
		
where $N_{10\text{B}}$ is the number of $^{10}$B atoms in the device ($n_{10\text{B}}$ multiplied by device volume $V$) and $A$ is the total area of the device. The device thickness has cancelled out of the equation ($V=Ad$). 

The boron neutron capture (BNC) process has an interaction cross section of $\sigma = 3840$\,barns for thermal neutrons, with an energy of $0.025$\,eV. The $0.22$\,mg of $96.6$\,\% $^{10}$B enriched B$_4$C added to sensitise the PNDI(2OD)2T to thermal neutrons resulted in a QE of $0.89$\,\% for these detectors. Using the middle of the estimated range given in Section~\ref{sec:Bcont} for the $^{10}$B content within the \textit{o}CbT$_2$-NDI sample, the QE of these detectors is estimated to be $0.69$\,\%.

The thermal neutron conversion efficiency ($\eta_{\text{n conv}}$) may be defined such that

\begin{equation}
    \eta_{\text{n conv}} =  \frac{|\Delta I_{th} |}{|\Delta I_{\alpha}|} \frac{\dot{N}_{\alpha}}{\dot{N}_{th}} \frac{E_{\alpha}}{E_{\text{BNC}}},
    \label{eq:etanconv}
\end{equation}

in which the thermal neutron irradiation tests are compared with benchmarking tests carried out with a sealed $370$\,kBq $^{241}$Am alpha radiation source at a standoff distance of $7$\,mm. Here, $\Delta I_{th}$ is the current enhancement due to thermal neutrons, and $\Delta I_{\alpha}$ is the current enhancement due to the $^{241}$Am alpha radiation. $\dot{N}_{\alpha}$ denotes the flux of alpha particles from the $^{241}$Am source and $E_\alpha$ is the alpha particle energy. $\dot{N}_{th}$ is the thermal neutron flux and $E_{\text{BNC}}$ is the energy given by the BNC process.

$\Delta I_{th}$ is found by the subtracting the unsensitised PNDI(2OD)2T current enhancement from the boron-containing samples’ current enhancements, removing current generated by the fast neutron component. $\dot{N}_{\alpha}$ is $62\,500$\,cm$^{-2}$s$^{-1}$ (from Taifakou et al.\cite{Taifakou:2021}), and $E_\alpha$ was calculated to be $3.6$\,MeV from calibration using an unencapsulated $3.7$\,kBq source under vacuum and in air measured by a commercial Cividec diamond detector rated with over $95$\,\% efficiency. For all calculations of $\eta_{\text{n conv}}$, $N_{th}$ has a value of $1.537$\,$\times$\,$10^7$\,cm$^{-2}$s$^{-1}$ with a $1.4$\,\% uncertainty given by standard beam calibration calculations provided by NPL. $E_{\text{BNC}}$ has a known literature value of $2.31$\,MeV which is the sum of the BNC process daughter particle kinetic energies. This value does not include the $0.482$\,MeV $\gamma$ photon given off in $94$\,\% of the interactions as these devices are not thick enough to interact significantly (less than $0.01$\,\% the radiation length). We obtain $\eta_{\text{n conv}}$ values of $\sim$$0.20$\,\% and $\sim$$0.14$\,\% for PNDI(2OD)2T:B$_4$C and \textit{o}CbT$_2$-NDI devices respectively at $+10$\,V bias.

\bigskip
\textbf{Copyright 2025 UK Ministry of Defence © Crown Owned Copyright 2025/AWE}


\balance


\end{document}